\documentclass{article}

\usepackage{amssymb}
\usepackage{amsmath}
\usepackage{amscd}
\newtheorem{lemma}{Lemma}

\begin{document}

\begin{titlepage}

\noindent
hep-th/9712182
\hfill ITP--UH--33/97 \\

\vskip 2.0cm

\begin{center}

{\large\bf CHIRAL BRST COHOMOLOGY OF N=2 STRINGS}\\

\medskip

{\large\bf AT ARBITRARY GHOST AND PICTURE NUMBER~$^*$}

\vskip 1.5cm

{\large Klaus J\"unemann \quad and \quad Olaf Lechtenfeld}

\vskip 0.5cm

{\it Institut f\"ur Theoretische Physik, Universit\"at Hannover}\\
{\it Appelstra\ss{}e 2, 30167 Hannover, Germany}\\
{E-mail: junemann, lechtenf@itp.uni-hannover.de}\\
 
\end{center}
\vskip 2.5cm

\begin{abstract}
We compute the BRST cohomology of the holomorphic part of the $N{=}2$ string 
at arbitrary ghost and picture number. We confirm the expectation that the 
relative cohomology at non-zero momentum consists of a single massless state 
in each picture. The absolute cohomology is obtained by an independent method
based on homological algebra. For vanishing momentum, the relative and absolute
cohomologies both display a picture dependence --- a phenomenon discovered 
recently also in the relative Ramond sector of $N{=}1$ strings by Berkovits and
Zwiebach~\cite{BZ}.
\end{abstract}

\vfill

\textwidth 6.5truein
\hrule width 5.cm
\vskip.1in

{\small \noindent ${}^{*\ {}}$
supported in part by the `Deutsche Forschungsgemeinschaft'; grant LE-838/5-1}

\end{titlepage}

\hfuzz=10pt

\section{Introduction}
The standard approach to describe quantum string theories is the BRST procedure 
which consists of introducing unphysical ghost fields associated with the 
symmetries of the theory. Physical states are then characterised as elements 
of the cohomology of the nilpotent BRST charge $Q$. 
For the open bosonic string this so-called {\it absolute} cohomology 
is well known to contain twice as many states as one would expect from 
light-cone quantisation~\cite{FGZ}. 
Each state appears in two copies -- either with or without the zero mode 
of the reparametrisation ghost $c_0$. The true physical spectrum therefore 
is determined by the BRST cohomology supplemented by the condition that 
a representative should be annihilated by the zero mode $b_0$ of the 
reparametrisation anti-ghost. 
This space defines the {\it relative} cohomology.\footnote{
For closed strings this kind of condition gets more complicated and leads 
to the concept of semi-relative cohomology. In this paper we consider 
for simplicity the chiral cohomology (describing open strings or the 
holomorphic part of closed strings) only.}

In the case of world-sheet supersymmetry, an additional subtlety arises 
due to the existence of an infinite number of inequivalent Fock space 
representations of the spinor ghosts -- the so-called picture degeneracy 
labelled by $\pi\in\frac12\mathbb{Z}$~\cite{FMS}.
In the $N{=}1$ string theory this problem is partly solved by bosonising 
the ghost fields, which allows one to construct a picture-raising operator~$X$ 
that maps physical states from the picture $\pi$ to $\pi{+}1$. Moreover, 
there exists a picture-lowering operator~$Y$ that inverts $X$ on the absolute
cohomology spaces, implying that the picture-raising operation is an isomorphism
of the cohomologies at different pictures~\cite{HMM}. Unfortunately, 
$Y$ does not commute with~$b_0$. 
Thus this argument does not guarantee that picture-raising is an isomorphism 
also of the {\it relative\/} cohomologies at different pictures. This problem 
has been addressed very recently by Berkovits and Zwiebach~\cite{BZ}, 
who used the momentum operator in the ${-}1$ picture to invert the zero mode 
of the picture-raising operator on states with {\it non-vanishing momentum\/}. 
This new picture-lowering operator commutes with $b_0$ and can therefore 
be used to prove the picture independence of the relative cohomology 
for non-vanishing momentum. 
However, these arguments do not rule out a picture dependence of the 
relative cohomology at {\it zero momentum\/} --- a phenomenon which indeed 
occurs in the R sector of the relative cohomology of $N{=}1$ strings~\cite{BZ}.

In $N{=}2$ string theory there exist two independent spinor ghost systems 
leading to two different picture numbers $(\pi^+,\pi^-)$. 
After bosonisation, one can construct picture-raising operators~$X^{\pm}$ 
in complete analogy to the $N{=}1$ case. These operators, however, 
cannot be inverted with local conformal fields~\cite{BKL,LP}. 
There is thus the immediate question whether or not the absolute or relative 
BRST cohomologies are identical at different pictures.

We address this question by two independent methods. 
The first method consists of applying the ideas of ref.~\cite{BZ} 
to the $N{=}2$ string. In contrast to conventional picture-lowering, 
this new kind of picture-lowering also works for the $N{=}2$ string, 
but only for non-vanishing momentum. 
Since it commutes with $b_0$, we confirm the picture independence of
both the absolute and the relative cohomology at non-zero momentum.

To describe the second method, let us recall that for the $N{=}1$ theory 
there exists an alternative argument, due to Narganes-Quijano~\cite{NQ}, 
that picture raising is an isomorphism. It makes use of the fact that 
bosonisation extends the Fock space by an additional oscillator and 
that in this extended space the absolute BRST cohomology is trivial. 
Some standard constructions from homological algebra then suffice to prove 
the isomorphy of the {\it absolute\/} cohomologies at different pictures. 
This work does not require the existence of an explicit 
picture-lowering operator and will be reviewed in more detail later on. 

For a specific choice of bosonisation, the absolute cohomology 
in the extended Fock space of the $N{=}2$ string again turns out to be trivial. 
However, the method of Narganes-Quijano cannot be carried over 
in a straightforward way, since the structure of the extended Fock space 
is more complicated for the $N{=}2$ string. We therefore need 
to slightly modify his method and invoke the spectral flow automorphism of the 
$N{=}2$ super Virasoro algebra~\cite{SS}. For the massless level,\footnote{
In principle, the proof works at any mass level. Its induction assumes the
equality of the absolute cohomology for some pair of neighbouring pictures,
which we only proved explicitly for the massless level.}
this will allow us to give an alternative proof of the picture independence 
of the {\it absolute\/} BRST cohomology at {\it non-vanishing momentum\/}. 
Unfortunately, we cannot treat the relative cohomology within this approach. 
It is, however, possible to extract some information about the exceptional case
of zero momentum.     

Nevertheless, most of our arguments fail for vanishing momentum, and 
we will demonstrate {\it picture dependence\/} of the exceptional cohomology
by explicit computations. For example, we shall see that the relative 
zero-momentum cohomology in the $({-}1,{-}1)$ picture consists of 
a single state of ghost number one, whereas in the $({-}1,0)$ picture 
there exist nontrivial states with any positive ghost number. 
In contrast to the $N{=}1$ string, this phenomenon occurs 
in the {\it absolute\/} cohomology as well, but it is possible to show that 
the picture dependence of the absolute zero-momentum cohomology 
is restricted to ghost numbers $0,1,2$ and $3$. We have checked that
this peculiar situation does not improve much when including 
the center-of-mass coordinate of the string~\cite{AB,BZ}.

The plan of the paper is as follows: In the next section we present 
a few basic facts about cohomology, clarify the relation between 
absolute and relative cohomology, and perform some explicit calculations 
in simple cases. Moreover, a complete computation of the cohomology 
in the $({-}1,{-}1)$ picture along the lines of refs.~\cite{FGZ,LZ}
and the role of spectral flow for the BRST cohomology are described.
In the third section we apply the ideas of Berkovits and Zwiebach~\cite{BZ} 
to the $N{=}2$ string and show that picture raising is a bijective map on 
both the absolute and relative cohomology classes for non-vanishing momentum.
In the fourth section we review part of the work of Narganes-Quijano and 
extend his method to the $N{=}2$ string to give an alternative treatment of 
picture raising. In the final section the results are summarised.

\section{Preliminary Investigations}
BRST quantisation and picture raising of the $N{=}2$ string has been reviewed 
recently in ref.~\cite{BL2} whose notation and conventions we adopt throughout
this paper. To keep things simple we concentrate on the NS sector. 
Other boundary conditions can be obtained by spectral flow~\cite{BL1,OV}.

\subsection{Relative and Absolute Cohomology} 
BRST-closed states with non-vanishing eigenvalues of the zero modes of the 
bosonic $N{=}2$ super Virasoro generators $L_0$ or $J_0$ are always exact. 
For cohomology computations it is therefore sufficient to restrict oneself 
to the space of  states that are annihilated by $L_0$ and $J_0$. 
Due to the relations
\begin{equation}
\{Q,b_0\} = L_0,
\hspace{1cm}
\{Q,\tilde{b}_0\} = J_0
\end{equation}
it is  possible to impose the further constraints that also the  
fermionic anti-ghost zero modes $b_0$ and $\tilde{b}_0$~\footnote{
As usual, $c$ and $b$ denote the reparametrisation ghosts. 
We write $\tilde{c}$ and $\tilde{b}$ for the $U(1)$ ghosts 
which have conformal weights $0$ and $1$, respectively.}
annihilate the states under consideration. This leads to the concept of 
{\it relative\/} cohomology which appears to have a more direct physical meaning
than the cohomology of the full Fock space. Throughout this paper we assume that
all states are annihilated by $\tilde{b}_0$ and thus work with the Fock space
\begin{equation}
F\ :=\ \{ |\psi \rangle \; ;\; L_0 |\psi \rangle = J_0 |\psi \rangle
= \tilde{b}_0 |\psi \rangle = 0 \}.
\end{equation}
The relative Fock space consists of states that are also annihilated by $b_0$:
\begin{equation}
F_{rel}\ :=\ \{ |\psi \rangle \in F \; ; \; b_0 |\psi \rangle = 0 \}.
\end{equation}
We treat the two types of fermionic anti-ghosts differently 
because it seems to be necessary to impose the conditions 
$J_0 |\psi\rangle = \tilde{b}_0 |\psi \rangle = 0$ 
as subsidiary conditions on an open $N{=}2$ string field 
in order to write down a free field action. The situation is quite similar to 
the field theory of closed bosonic strings where the conditions 
$(L_0 - \bar{L}_0) |\psi\rangle = (b_0 - \bar{b}_0 ) |\psi\rangle = 0$ 
have to be imposed~\cite{Z}. In contrast, $b_0 |\psi \rangle = 0$ 
can be considered as a gauge-fixing condition (Siegel gauge), 
and $L_0 |\psi\rangle = 0$ simply is the equation of motion.

Both the spaces $F$ and $F_{rel}$ possess a grading 
with respect to picture  and  ghost number:
\begin{equation}
F = \sum_{g,\pi^+, \pi^-} F^{g,\pi^+, \pi^-},
\hspace{1cm} 
F_{rel} = \sum_{g,\pi^+, \pi^-} F_{rel}^{g,\pi^+, \pi^-}.
\end{equation}
We often suppress the obvious grading with respect to the
center-of-mass momentum $k\in\mathbb{R}^{2,2}$.
Following ref.~\cite{BKL} we bosonise the (commuting) spinor ghosts,
\begin{equation}
\gamma^{\pm} \to \eta^{\pm} e^{\varphi^{\pm}},
\hspace{1cm}
\beta^{\mp}  \to e^{-\varphi^{\pm}} \partial \xi^{\mp},
\end{equation}
and define the (total) ghost number current in a slightly unsual way~\cite{B}:
\begin{equation}
j_{gh}\ =\ - b c - \tilde{b} \tilde{c} + \eta^+ \xi^- + \eta^- \xi^+.
\end{equation}
This has the advantage of commuting with picture raising 
while still assigning the correct ghost number to all ghost fields
and giving $\xi^\pm$ the ghost number minus one.
Moreover, we define the ghost number of the ground state 
in the $(0,0)$ picture (and therefore in all pictures) to be zero.
The BRST cohomology spaces inherit the various gradings and are denoted by
\begin{equation}
H(F) = \sum_{g,\pi^+, \pi^-} H^{g,\pi^+, \pi^-}(F),
\hspace{1cm} 
H(F_{rel}) = \sum_{g,\pi^+,\pi^-} H ^{g,\pi^+, \pi^-}(F_{rel}).
\end{equation}
$H(F)$ is called the {\it absolute\/}\footnote{
Obviously, this name in not entirely logical since our absolute cohomology 
is still relative with respect to $\tilde{b}_0$. 
The relation between $H(F)$  and the cohomology of the full Fock space 
(where also the $\tilde{b}_0$ condition is relaxed) 
can be analysed straightforwardly by the methods described in this section and 
is not relevant to the picture degeneracy which is the subject of this paper.} 
and $H(F_{rel})$ the {\it relative\/} cohomology. These two types of cohomology 
are related by a well known exact sequence~\cite{FGZ}. 
Although this has been described in detail in refs.~\cite{LZ,WZ},
we briefly repeat this analysis here.

$F$ and $F_{rel}$ differ just by the possibility to apply the oscillator $c_0$,
which implies the decomposition $F = F_{rel} \oplus c_0 F_{rel}$.  The 
inclusion $i:F_{rel} \to F$ and the projection $pr: F \to F_{rel}$, defined as 
\begin{equation}
i (\psi) := \psi + c_0 0,
\hspace{1cm}
pr (\psi + c_0 \chi) := \chi,
\hspace{1cm}
\psi\; ,\chi \in F_{rel},
\end{equation}
can be combined to the following exact sequence:
\begin{equation}\label{s1}
0 \longrightarrow F_{rel} \stackrel{i}{\longrightarrow} F  
\stackrel{pr}{\longrightarrow} F_{rel}  \longrightarrow 0.
\end{equation}
Since the inclusion and the projection both commute with the BRST operator~$Q$,
this exact sequence induces an exact cohomology triangle:\footnote{
This is a standard mathematical construction; 
see for example chapter 0 of ref.~\cite{GHV} for a review.}

\begin{center}
\unitlength0.5cm
\begin{picture}(9,4)
\put(0,0){$H(F_{rel})$}
\put(1.2,1){\vector(1,1){2}}
\put(1.7,2.2){$i$}
\put(3.3,3.2){H(F)}
\put(4.5,3){\vector(1,-1){2}}
\put(5.7,2.2){$pr$}
\put(6,0){$H(F_{rel})$}
\put(5.6,0.2){\vector(-1,0){3}}
\put(3.3,0.5){$\{Q,c_0\}$}
\end{picture}
\end{center}

The connecting homomorphism carries ghost number 2 and thus allows us 
to unwind the above triangle into the long exact cohomology sequence
\begin{equation}\label{seq1}
 \longrightarrow H^{g+1}(F)  \stackrel{pr}{\longrightarrow} 
H^g(F_{rel})  \stackrel{\{Q,c_0\}} {\longrightarrow} H^{g+2}(F_{rel})  
\stackrel{i}\longrightarrow H^{g+2}(F) \stackrel{pr}{\longrightarrow} 
H^{g+1}(F_{rel})  \longrightarrow. 
\end{equation}
This sequence will turn out to be useful for explicit calculations. 
It is interesting that picture raising can be treated  similarly~\cite{NQ}  
as we will show in section 4.

\subsection{Explicit Computations in the Massless Sector}
The simplest possible case for explicit computations is the massless sector  
in the $({-}1,{-}1)$ picture where all positively moded spinor ghost 
oscillators are annihilation operators.
The relative Fock space $F_{rel}^{-1,-1}(k{\cdot}k{=}0)$ consists  
of a single state with ghost number $g=1$, namely
\begin{equation} \label{state1}
c_1 |\; {-}1,{-}1 , k \;\rangle, \hspace{1cm} k{\cdot}k = 0,
\end{equation}
where $|\;\pi^+,\pi^-  , k \;\rangle$ denotes the ground state 
with momentum $k$ in the $(\pi^+,\pi^-)$ picture.
The state (\ref{state1}) is BRST invariant but not exact and thus constitutes  
the complete relative cohomology in the $({-}1,{-}1)$ picture. 
This is the analogue of the vanishing theorems in the massless sector 
for the bosonic~\cite{FGZ} and the $N{=}1$ string~\cite{LZ}.
The sequence (\ref{seq1}) implies that the absolute cohomology contains 
two states: the state given in (\ref{state1}) and 
\begin{equation} \label{state2}
c_0 c_1 |\; {-}1,{-}1 , k \;\rangle , \hspace{1cm} k{\cdot}k = 0 .
\end{equation} 
The corresponding vertex operators creating these states from the 
$(0,0)$ picture vacuum are  
\begin{equation}\label{v-1-1}
V^{(1)}_{({-}1,{-}1)} (z)  = c e^{- \varphi^+ - \varphi^-} e^{ik\cdot Z} (z),
\hspace{1cm}
V^{(2)}_{({-}1,{-}1)} (z)  = c \partial c e^{- \varphi^+ - \varphi^-} 
e^{ik\cdot Z} (z).
\end{equation}
We will see shortly that the connection between the relative and the absolute 
cohomology is more complicated in other pictures, since multiplying 
a state from the relative cohomology by $c_0$ does not in general produce a 
BRST-closed state. In the $({-}1,{-}1)$ picture everything carries over 
unchanged to the exceptional case $k{=}0$, i.e.
\begin{equation}
H^{-1,-1}(k{=}0)\ =\ H^{-1,-1}(k{\cdot}k{=}0) 
\qquad{\rm for}\ F\ {\rm and}\ F_{rel}.
\end{equation}

We now turn to the massless sector of the $({-}1,0)$ picture 
where $\gamma^+_{1/2}$ becomes a creation operator.  
The relative Fock space $F_{rel}^{-1,0}(k{\cdot}k{=}0)$ 
is spanned by the following states with ghost number~$g$: 
\begin{align}\label{states}
A_N^{\mu}\ &:=\ c_1 (\gamma^+_{1/2})^N (\gamma^-_{-1/2})^N d_{-1/2}^{-\mu}
|\; {-}1,0, k \;\rangle,&  g& = 2N+1  \nonumber \\
B_N\ &:=\ c_1 (\gamma^+_{1/2})^N  (\gamma^-_{-1/2})^{N+1} 
|\; {-}1 , 0, k \;\rangle,&    g& = 2N+2 \\
C_N^{\mu\nu}\ &:=\ c_1 (\gamma^+_{1/2})^{N+1}(\gamma^-_{-1/2})^N d_{-1/2}^{-\mu}
d_{-1/2}^{-\nu}|\; {-}1 ,  0, k \;\rangle,&    g &= 2N+2\nonumber
\end{align}
where $N$ is a non-negative integer, $\mu = 0, 1$,  and 
\begin{equation}
\gamma_r^{\pm} = \oint \frac{dz}{2\pi i} z^{r-3/2} \gamma^{\pm} (z), 
\hspace{1cm}
d_r^{\pm\mu} = \oint \frac{dz}{2\pi i} z^{r-1/2}  i \psi^{\pm\mu} (z) 
\end{equation}
are the Fourier modes of the spinor ghosts and matter fermions.
The BRST operator acts as
\begin{eqnarray}
Q A_N^{\mu} &=& 2 k^{-\mu} B_N + k_{\nu}^+  C_N^{\nu\mu} \nonumber \\
Q B_N &=&  k_{\mu}^+ A_{N+1}^{\mu} \\
Q C_N^{\mu\nu} &=& 2 k^{-\mu} A_{N+1}^{\nu} - 2 k^{-\nu} A_{N+1}^{\mu}\nonumber
\end{eqnarray}
($Q^2 = 0$ can be checked explicitly). By inspection one learns that 
the cohomology $H^{-1,0}(F_{rel}|k{\cdot}k{=}0)$ resides at $g=1$ only 
and is represented by
\begin{equation}\label{rep}
k^+_{\mu} A_0^{\mu}\ =\ c_1 k^+ {\cdot} d^-_{-1/2} |\; {-}1 , 0, k \;\rangle,
\hspace{1cm} k{\cdot} k = 0
\end{equation}
for any non-vanishing value of the momentum.\footnote{
Note that in our conventions $k^+$ and $k^-$ are related by complex conjugation
and thus cannot vanish individually.
This is different in a real $SL(2,\mathbb{R})$ notation~\cite{LS,DL},
where $k^+=0$ is possible with non-zero~$k^-$.
In such a case the representative (\ref{rep}) can be replaced by
$\epsilon_{\mu\nu} k^{-\nu} A_0^{\mu}$, but the cohomology is unchanged.}

The corresponding vertex operator creating this state from the 
$(0,0)$ picture vacuum is
\begin{equation}
V^{(1)}_{({-}1,0)}(z)  = 
c k^+ {\cdot} \psi^- e^{-\varphi^-} e^{i k {\cdot} Z} (z)
\end{equation}
which is the picture-raised version of $V^{(1)}_{({-}1,{-}1)}$ in (\ref{v-1-1}) 
(see the appendix of ref.~\cite{BKL} for a detailed list of vertex operators).
This proves that in this simple case the picture-raising operation $X^-$
(and similarly $X^+$) is an isomorphism between the relative cohomologies
at $k{\neq}0$.

What about the absolute cohomology $H^{-1,0}(F|k{\cdot}k{=}0)$? 
The sequence~($\ref{seq1}$) implies that it is non-vanishing 
only at ghost number one and two. Obviously, 
the ghost number one part is simply represented by $k^+_{\mu} A_0^{\mu}$. 
Applying $Q$ to $c_0 k^+_{\mu} A_0^{\mu}$ yields
\begin{equation} 
Q c_0 k^+_{\mu} A_0^{\mu}\ =\ -4 k^+_{\mu} A_1^{\mu}\ =\ - 4 Q B_0,
\end{equation}
showing that the cohomology class at ghost number two is represented by 
$ c_0 k^+_{\mu} A_0^{\mu} + 4 B_0$.
The two corresponding vertex operators are $V^{(1)}_{({-}1,0)}$ and
\begin{equation}
V^{(2)}_{({-}1,0)}(z) = 
\big{(}  c \partial c k^+ {\cdot} i\psi^- e^{-\varphi^-} + 
4  c \eta^- \big{)}e^{i k \cdot Z} (z)
\end{equation}
which are both obtained by picture raising the vertex operators 
in~(\ref{v-1-1}). For non-zero momentum we thus see that 
picture raising is an isomorphism in the absolute cohomology, too.
Together, we have 
\begin{equation}\label{start}
X^-:\quad H^{-1,-1}(k{\cdot}k{=}0)\ 
\stackrel{\cong}{\longrightarrow}\
H^{-1,0}(k{\cdot}k{=}0)
\qquad {\rm at}\quad k\neq0
\end{equation}
for $F$ as well as for $F_{rel}$.

In the exceptional case, $k=0$, things are strikingly different. 
$Q$ vanishes identically on the relative Fock space $F_{rel}^{-1,0}(k{=}0)$, 
and any of the states in~(\ref{states}) represents 
its own nontrivial cohomology class even though 
the picture-raising operation annihilates the $({-}1,{-}1)$ vertex operator. 
Moreover, explicit calculations at higher pictures seem to indicate 
a proliferation of physical states. Therefore, 
the exceptional relative cohomologies $H^{\pi^+,\pi^-}(F_{rel}|k{=}0)$ 
look entirely different in various pictures.

To work out the exceptional absolute cohomology, 
we additionally have to consider the states in (\ref{states}) 
multiplied by $c_0$. For $k{=}0$ one finds that $Q$ acts on these states as
\begin{eqnarray}
Q c_0 A_N^{\mu} &=& -4 A_{N+1}^{\mu} \nonumber \\
Q c_0 B_N &=&  -4 B_{N+1} \\
Q c_0 C_N^{\mu\nu} &=& -4 C_{N+1}^{\mu\nu} \nonumber.
\end{eqnarray}
Obviously, the absolute zero-momentum cohomology $H^{-1,0}(F|k{=}0)$ 
is spanned by the two states $A_0^{\mu}$ at ghost number one, 
by two more, $B_0$ and $C_0^{\mu\nu}=-C_0^{\nu\mu}$, at ghost number two, 
and vanishes at any other ghost number. 

Are these results for $H^{-1,0}(k{=}0)$ consistent 
with the sequence~(\ref{seq1})? 
At odd positve ghost number $g=2N+1$, the relative cohomology is spanned  by 
$A_N^{\mu}$ which contains $N$ powers of $\gamma^+_{1/2} \gamma^-_{-1/2}$. 
The connecting homomorphism $\{Q,c_0\}$ acts (up to a numerical factor) 
by multiplication of just such a factor. We thus see that it is an isomorphism 
between the relative cohomologies with odd ghost number. The same is true 
for positive even ghost number. But this is precisely what we learn from 
the sequence (\ref{seq1}) if we insert the result that the absolute cohomology 
vanishes at ghost number greater than 2.

Let us briefly summarise the result of the above calculations 
for $k{\cdot}k{=}0$: 
At non-zero momentum, picture raising establishes an isomorphy 
between the $({-}1,{-}1)$ and the $({-}1,0)$ pictures 
for both the relative and the absolute cohomology. 
For vanishing momentum, however, the cohomologies look very different. 
In the $({-}1,{-}1)$ picture both the absolute and the relative cohomology are 
obtained by the zero-momentum limit of the cohomology at non-vanishing momentum.
In the $({-}1,0)$ picture the BRST operator vanishes in the relative Fock space.
The relative cohomology is two-dimensional at any positive ghost number. 
In contrast, the absolute cohomology is two-dimensional only at ghost numbers 
one and two and vanishes elsewhere.

In other pictures one finds non-trivial absolute cohomology classes also 
at ghost numbers zero and three. For example, the states 
\begin{equation}
|  0,0, k{=}0 \rangle 
\hspace{1cm} {\rm and} \hspace{1cm}
c_{-1} c_0 c_1 |  -2,-2, k{=}0 \rangle 
\end{equation} 
are both BRST invariant but not exact. 
This is in contrast to the $N{=}1$ string where picture-lowering guarantees the
picture independence of the absolute cohomology even in the exceptional case.
In section~4, however, we will prove that the absolute exceptional cohomology 
vanishes for ghost number $g\neq 0, 1, 2, 3$ at any picture. 
The picture dependence can thus only occur for these ghost numbers.

\subsection{Complete Calculation in the $({-}1,{-}1)$ Picture}
The above calculations were all done for $k{\cdot}k=0$. 
But what about massive states? Surely such states would carry 
additional Lorentz indices and therefore  describe higher spin fields. 
Due to the absence of transverse dimensions in the (2,2) space-time, 
these states should not contribute any physical degrees of freedom, 
leaving the ground state as the only physical state.
Although this sounds very plausible it is not what one would call 
a rigorous computation of the relative BRST cohomology. 
The most powerful approach  to this kind of problem has been invented 
by Frenkel, Garland and Zuckerman~\cite{FGZ} and extended to the 
$N{=}1$ string in the ${-}1$ picture by Lian and Zuckerman~\cite{LZ}. 
Their method consists of introducing a new kind of grading -- 
the filtration degree -- to reduce the computation of the BRST cohomology 
to a standard problem of Lie algebra cohomology and can be applied 
to the $N{=}2$ string, as well. Its essential new feature, 
namely the existence of the additional bosonic current~$J$, 
can be incorporated in a straightforward way by simply extending 
the definition of the relative Fock space as indicated in section 2.1. 
Another important ingredient in this analysis is that the Fock space 
of the matter sector must be a free module of the algebra 
of the negatively moded $N{=}2$ super Virasoro generators. 
This property is also satisfied for critical $N{=}2$ strings. 
For non-vanishing momentum it has in fact been shown in ref.~\cite{Bi} that
the Fock space is a direct sum of universal enveloping algebras 
of the negative $N{=}2$ super Virasoro algebra.\footnote{
For $k{=}0$ this is not true since the ground state is then annihilated 
by $L_{-1}$. As in other string theories, for this reason 
such kind of analysis does not apply in the exceptional case.}
The rest of the argument works in complete analogy to the $N{=}1$ string, 
and it does not seem necessary to repeat it here since it has been described 
in great detail in ref.~\cite{LZ}.
One finally arrives at the expected result that the state (\ref{state1}) 
is the only physical degree of freedom in the $({-}1,{-}1)$ picture 
and that there is no room for discrete states or other surprises. 
For $k{\cdot} k > 0$ we thus have
\begin{equation}\label{FGZ}
H^{g,-1,-1}(F)\ =\ H^{g,-1,-1}(F_{rel})\ = 0 
\hspace{1cm} {\rm for\; any}\; g.
\end{equation}
Unfortunately, this kind of analysis applies only to the $({-}1,{-}1)$ picture.
The latter is singled out as the only picture 
where the creation (annihilation) operators are precisely the 
negatively (positively) moded oscillators and which has a nondegenerate scalar 
product with itself. Perhaps it is possible to find a clever redefinition 
of the filtration degree to apply this method also to other pictures, 
but it is not obvious to the authors how this could be done.

\subsection{Spectral Flow}
We finally discuss one further aspect of the $N{=}2$ string, 
namely spectral flow~\cite{SS,BL1}. 
However, this will only be needed  for the discussion in section 4.

Spectral flow is an automorphism of the $N{=}2$ superconformal algebra 
associated to the $U(1)$ subalgebra. 
An explicit construction is presented in the appendix of ref.~\cite{BL2}. 
If the spectral flow parameter $\Theta$ is chosen from the interval $(0,1)$, 
the spectral flow operator ${\cal S} (\Theta )$  relates sectors 
with different boundary conditions 
(see however ref.~\cite{LP} for a different point of view). 
For $\Theta = 1$ it is a map within each sector
and has a number of useful properties~\cite{BL1}: 
it has zero ghost number, commutes with $Q$, changes $\pi^+$ by $+1$, 
$\pi^-$ by $-1$ and is invertible (choose $\Theta = -1$). 
It is therefore an isomorphism of the cohomologies,
\begin{equation}
{\cal S}(1):\quad H^{\pi^+,\pi^-}(F)\ \stackrel{\cong}{\longrightarrow}\ 
H^{\pi^++1,\pi^--1}(F),
\end{equation}
and it follows by induction that 
\begin{equation}\label{SFO1}
H^{\pi^+,\pi^-}(F)\ \cong\ H^{\pi^++n,\pi^--n}(F)
\end{equation}
for arbitrary $\pi^+,\pi^-,k$ and any integer $n$.
Moreover, ${\cal S} (1)$ commutes with the picture-raising operators~$X^{\pm}$
up to BRST trivial terms~\cite{BL1}, 
i.e. it commutes with them on the cohomology spaces. 

Since we have seen above that, for non-vanishing momentum, $X^-$ is an 
isomorphism between $H^{-1,-1}(F|k{\cdot}k{=}0)$ and 
$H^{-1,0}(F|k{\cdot}k{=}0)$, the commutative diagram
$$
\begin{CD}
H^{-1,-1}(F|k{\cdot}k{=}0) @>{X^-}>{\cong}> H^{-1,0}(F|k{\cdot}k{=}0)\\
@V{{\cal S}(1)^n}V{\cong}V @V{{\cal S}(1)^n}V{\cong}V \\
H^{-1+n,-1-n}(F|k{\cdot}k{=}0) @>{X^-}>> H^{-1+n,n}(F|k{\cdot}k{=}0)
\end{CD}
$$
implies that $X^-$ is also an isomorphism in the bottom row.
Thus, the spaces
\begin{equation}
H^{\pi^+,\pi^-}(F|k{\cdot}k{=}0) \qquad {\rm for} \quad
\pi^+{+}\pi^-\in\{-2,-1\}
\end{equation}
are all isomorphic for non-zero momentum.
Finally, let us remark that the above argument is not true 
for the relative cohomology since ${\cal S} (1)$ does not commute with $b_0$.

\section{Picture-Lowering}
In this section we apply the method of section 2 of ref.~\cite{BZ} 
to the open $N{=}2$ string. We will, however, refrain from presenting 
the details since the calculations carry over in a straightforward way. 

To begin with, let us recall the bosonisation of the spinor ghosts 
of the $N{=}2$ string~\cite{BKL}:
\begin{equation}
\gamma^{\pm} (z) \to \eta^{\pm} e^{\varphi^{\pm}} (z), 
\hspace{1cm}
\beta^{\mp} (z) \to e^{-\varphi^{\pm}} \partial \xi^{\mp}.
\end{equation}
The zero modes $\xi_0^{\pm}$ of the weight-zero fields $\xi^{\pm}(z)$ 
do not take part in this process, and thus the Fock space $F$ 
is extended to the bigger space $\bar{F}$. 
The picture-raising operators acting on $F$ are defined as 
\begin{equation}
X^{\pm}_0 := \{ Q , \xi^{\pm}_0 \} = \oint \frac{dz}{2\pi i z} X^{\pm} (z),
\hspace{1cm} X^{\pm} (z) := \{Q,\xi^{\pm} (z)\}
\end{equation}
and map a BRST-closed state $|\psi \rangle \in F $ to 
$Q \xi^{\pm}_0 |\psi \rangle $ which is trivial in $\bar{F}$ but not in $F$. 
Note that both $X^{\pm}_0$ do not contain any $\xi^{\pm}_0$ 
and therefore are maps within the small space $F$. 

Following ref.~\cite{BZ} we consider the momentum operators 
in the $({-}1,0)$ and $(0,{-}1)$ picture:
\begin{equation}
\tilde{p}^{\pm\mu}\ =\ 
\oint\frac{dz}{2\pi i}\;e^{-\varphi^{\pm}}i\psi^{\pm\mu}. 
\end{equation}
Because of 
\begin{equation}\label{bew0}
[ Q , e^{-\varphi^{\pm}} i\psi^{\pm\mu}  ]\ =\ 
\partial ( c e^{-\varphi^{\pm}} i\psi^{\pm\mu} ),
\end{equation}
$\tilde {p}^{\pm\mu}$ is BRST invariant and satisfies the key relations
\begin{equation}\label{key}
X^{\pm}_0 \tilde {p}^{\mp\mu}\ =\ 2p^{\mp\mu} + \{Q , m^{\pm\mu}\},
\hspace{1cm}
\tilde {p}^{\mp\mu} X^{\pm}_0\ =\ 2p^{\mp\mu} + \{Q , n^{\pm\mu}\},
\end{equation}
where $p^{\pm\mu}$ is the center-of-mass momentum,
\begin{equation}
p^{\pm\mu}\ =\ \oint \frac{dz}{2\pi i }\; i\partial Z^{\pm\mu},
\end{equation}
and $m^{\pm\mu}$ and $n^{\pm\mu}$ are given by
\begin{equation}
m^{\pm\mu}\ =\ \oint \frac{dz_1}{2\pi i z_1 } \oint_{|z_2| < |z_1|}  
\frac{dz_2}{2\pi i } \int_{z_2}^{z_1}\!\! dw\; 
\partial \xi^{\pm} (w)\ e^{-\varphi^{\mp}} i\psi^{\mp\mu} (z_2), 
\end{equation}
\begin{equation}
n^{\pm\mu}\ =\ \oint  \frac{dz_2}{2\pi i } \oint_{|z_1| < |z_2|}  
\frac{dz_1}{2\pi i z_1 }\int_{z_2}^{z_1}\!\! dw\;  
e^{-\varphi^{\mp}} i\psi^{\mp\mu} (z_2)\ \partial \xi^{\pm} (w). 
\end{equation}
The proof of the analogue of equations (\ref{key}) for the $N{=}1$ string has
been given in ref.~\cite{BZ}, section 2, and works in our present case, as well. 

For completeness we present the calculation that establishes 
the first of equations (\ref{key}). In terms of conformal fields 
the expression $X^+_0 \tilde {p}^{-\mu}$ reads
\begin{equation}\label{bew1}
X^+_0\tilde{p}^{-\mu}\ =\ \oint\frac{dz_1}{2\pi i z_1 }\oint_{|z_2|<|z_1|}
\frac{dz_2}{2\pi i }\; X^+ (z_1)\; e^{-\varphi^-} i\psi^{-\mu} (z_2).
\end{equation}
As the fields $X^+$ and $e^{-\varphi^-} i\psi^{-\mu}$ approach each other, 
no singularity appears since they have the short distance expansion 
\begin{equation}
X^+ (z)\; e^{-\varphi^-} i\psi^{-\mu} (w)\ \sim\ 
c\partial \xi^+ e^{-\varphi^-} i\psi^{-\mu} (w)\
+\ 2 i \partial Z^{-\mu} (w)\ +\ {\cal O} (z-w).
\end{equation}
We can therefore insert the relation
\begin{equation}
X^+ (z_1)\ =\ X^+(z_2)\ +\ \int_{z_2}^{z_1}\!\! dw\; \{ Q,\partial \xi^+ (w)\}
\end{equation}
into (\ref{bew1}) and obtain
\begin{eqnarray}\label{bew2}
X^+_0 \tilde {p}^{-\mu} &=& 
\oint\frac{dz_2}{2\pi i}\; (c\partial \xi^+ e^{-\varphi^-} i\psi^{-\mu}
+ 2 i \partial Z^{-\mu} )(z_2) \\
&&+\  \oint \frac{dz_1}{2\pi i z_1 } \oint_{|z_2| < |z_1|}  
\frac{dz_2}{2\pi i } \int_{z_2}^{z_1}\!\! dw \; 
\{ Q , \partial \xi^+ (w) \}\ e^{-\varphi^-} i\psi^{-\mu} (z_2). \nonumber
\end{eqnarray}
With the help of equation  (\ref{bew0}) 
the integrand of the last term can be rewritten as 
\begin{eqnarray}
\{ Q , \partial \xi^+ (w) \} \ e^{-\varphi^-} i\psi^{-\mu} (z_2) &=& 
\{ Q , \partial \xi^+ (w)  e^{-\varphi^-} i\psi^{-\mu} (z_2) \} \nonumber \\
&&+\ \partial \xi^+ (w)\ \partial ( c e^{-\varphi^-} i\psi^{-\mu} (z_2)).
\end{eqnarray}
The second integral in (\ref{bew2}) thus becomes
\begin{eqnarray}
&&\{ Q\ ,\ \oint \frac{dz_1}{2\pi i z_1 } \oint_{|z_2| < |z_1|}  
\frac{dz_2}{2\pi i } \int_{z_2}^{z_1}\!\! dw\; 
\partial \xi^+ (w)\ e^{-\varphi^-} i\psi^{-\mu} (z_2) \}  \nonumber \\
&+& \oint \frac{dz_1}{2\pi i z_1 } \oint_{|z_2| < |z_1|}  
\frac{dz_2}{2\pi i }\;  ( \xi^+ (z_1) - \xi^+(z_2) )\ 
\partial ( c e^{-\varphi^-} i\psi^{-\mu} (z_2))   \nonumber \\
&=& \{Q , m^{+\mu}\} - \oint   \frac{dz_2}{2\pi i }\;                 
\xi^+(z_2) \ \partial ( c e^{-\varphi^-} i\psi^{-\mu} (z_2)).
\end{eqnarray}
Substituting this back into (\ref{bew2}) yields
\begin{equation}
X^+_0 \tilde {p}^{-\mu} = 
2p^{-\mu} + \{Q , m^{+\mu}\} + \oint\!\frac{dz_2}{2\pi i } \; 
( c\partial \xi^+ e^{-\varphi^-} i\psi^{-\mu} - \xi^+   
\partial ( c e^{-\varphi^-} i\psi^{-\mu}))(z_2). 
\end{equation}
This proves the first of equations (\ref{key}) since the integrand 
in the last term is a total derivative and the integral thus vanishes.

Because $p^{\pm\mu}$ is picture-neutral,
the relations (\ref{key}) ensure that $X^{\pm}_0$ are bijective maps 
between absolute cohomology classes at non-zero momentum and therefore 
prove their picture independence (see ref.~\cite{BZ} for more details). 
Moreover, all operators involved commute with $b_0$ and $\tilde{b}_0$ and 
thus generalise the results to the relative cohomologies.
Although obvious, let us emphasise that the above argument is invalid
on states with vanishing momentum. There is no contradiction 
to the results of section 2.2.

\section{An Alternative Proof} 
In this section we give an alternative proof, inspired by ref.~\cite{NQ}, 
that picture raising is an isomorphism of the {\it absolute massless\/} 
cohomology for non-vanishing momentum. 
Here we do not refer to any kind of picture-lowering, 
and thus this analysis is logically independent from that of section 3. 
After all, it is good to have two seperate proofs of one statement. 
Unfortunately, we can only treat the absolute, but not the relative cohomology
within this approach.
We are able, however, to obtain some information about the picture dependence 
of the absolute cohomology in the exceptional ($k{=}0$) case.

Before considering the cohomology of the $N{=}2$ string at arbitrary picture, 
let us briefly review part of the work of Narganes-Quijano~\cite{NQ}.

\subsection{The $N{=}1$ String}
Bosonisation in the $N{=}1$ theory consists of replacing the  
$\gamma$ and $\beta$ ghosts by~\cite{FMS} 
\begin{equation}
\gamma \rightarrow \eta e^{\varphi},
\hspace{1cm}
\beta \rightarrow e^{-\varphi} \partial \xi.
\end{equation}
As already mentioned in section 3, this extends the Fock space $F$ 
to the larger space  $\bar{F} = F \oplus \xi_0 F$.
We thus have a situation completely analogous to that described in section 2.1.
Consider the inclusion $i: F \mapsto \bar{F}$ 
and the projection $pr: \bar{F} \mapsto F$, defined as
\begin{equation}
i(a) := a + \xi_0 0,
\hspace{1cm} 
pr (a  + \xi_0 b ) := b,
\hspace{1cm} 
a ,  \; b \in F.
\end{equation}
Note that the projection has ghost number one and picture number minus one.
The corresponding exact sequence is
\begin{equation}\label{seq2}
0 \longrightarrow F \stackrel{i}{\longrightarrow} \bar{F}  
\stackrel{pr}{\longrightarrow} F  \longrightarrow 0.
\end{equation}
Since both the inclusion and the projection (anti-)commute with $Q$, 
this exact sequence again induces an exact cohomology triangle. 
The connecting homomorphism here is nothing but the picture-raising operator 
$X_0 = \{Q, \xi_0 \}$! Including the gradation with respect to picture number 
yields the long exact sequence
\begin{equation}
\ldots \longrightarrow H^{\pi}(\bar{F}) \stackrel{pr }{\longrightarrow} 
H^{\pi - 1}(F) \stackrel{X_0}{\longrightarrow} H^{\pi}(F)  
\stackrel{i}\longrightarrow H^{\pi}(\bar{F})  \longrightarrow \ldots
\end{equation}
between the various cohomology spaces.

Now, the key observation is that the BRST cohomology of $\bar{F}$ is trivial! 
This follows immediately from the existence of the operator 
$W = 4 c \xi \partial \xi e^{-2\varphi} $ with the property 
\begin{equation}
\{ Q , W(z) \} = 1.
\end{equation}
Inserting  $H^{\pi}(\bar{F}) = 0 $ for arbitrary $\pi$ into the above sequence 
implies that $X_0$ is an isomorphism between $H^{\pi-1}(F)$ and $H^{\pi}(F)$, 
without refering to any kind of  picture-lowering  operator.
It is, however, important to note that the operator $W$ 
does not commute with $b_0$. Therefore, the cohomology 
in the large relative space $\bar{F} \cap {\rm Ker} b_0$ need not be trivial. 
Correspondingly, the above construction does not imply that the picture-raising  
operation is an isomorphism between the relative cohomologies, as well.

\subsection{The $N{=}2$ String}
In $N{=}2$ string theory, bosonisation extends the Fock space 
by {\it two\/} additional oscillators:
\begin{equation}
\bar{F}\ =\ F \oplus \xi_0^+ F \oplus \xi_0^- F \oplus \xi_0^+ \xi^-_0 F.
\end{equation}
The first step is to check whether the cohomology in the large space is trivial.
Indeed, there exists an operator $W$ with the right property, namely\footnote{
There is a misprint in eqn.~(7.3) of ref.~\cite{BL2}. 
The second $\tilde{b}$ must be replaced by $b$; 
it is this term that produces a pole when contracted with $W$.}
\begin{equation}\label{W}
W (z) = - \frac{1}{4} c \xi^+ \xi^- e^{\varphi^+} e^{\varphi^-} (z),
\hspace{2cm}
[Q, W(z)] = 1.
\end{equation}
This result only holds for this special choice of bosonisation. 
If one bosonises a different linear combination of the spinor ghosts, 
the corresponding $W$ does not exist. 
For the cohomology in the small space this is however irrelevant.
As in the $N{=}1$ theory, the operator $W$ does not commute with $b_0$ so that 
we cannot obtain information about the relative cohomology within this approach.

The situation is more complicated than for the $N{=}1$ string, 
because the small and the large Fock space cannot be connected 
in such a simple way as in (\ref{seq2}). We thus have to proceed in two steps. 
First let us define
\begin{equation}
F_{\pm}\ :=\ F \oplus \xi^{\pm}_0 F
\end{equation}
and the projection 
\begin{equation}
pr: \bar{F} \to F_-
\qquad{\rm by}\qquad
pr (a + \xi^-_0 b + \xi^+_0 c + \xi^+_0 \xi^-_0 d ) := c + \xi^-_0 d
\end{equation}
for $a$, $b$, $c$,  $d \in F$. 
The map $pr$ again has picture number minus one and anticommutes with $Q$ since
\begin{eqnarray}
&&(pr \circ Q) (a + \xi^-_0 b + \xi^+_0 c + \xi^+_0 \xi^-_0 d) \nonumber \\
&&\quad =\ pr (Qa + Q\xi^-_0 b + Q\xi^+_0 c + Q\xi^+_0 \xi^-_0 d)\nonumber\\
&&\quad =\ pr (Qa + X_0^- b - \xi^-_0 Qb + X^+_0 c - \xi^+_0 Qc 
+ X_0^+ \xi^-_0 d - \xi^+_0 Q \xi^-_0 d) \nonumber \\
&&\quad =\ - Q ( c + \xi^-_0 d) \nonumber \\
&&\quad =\ - (Q \circ pr) (a + \xi^-_0 b + \xi^+_0 c + \xi^+_0 \xi^-_0 d) 
\end{eqnarray}
where all $\xi^{\pm}_0$ are explicit.
Together with the inclusion  $i: F_- \mapsto \bar{F}$ 
(which trivially commutes with Q) one can form the exact sequence 
\begin{equation}
0 \longrightarrow F_- \stackrel{i}{\longrightarrow} \bar{F}  
\stackrel{pr}{\longrightarrow} F_-  \longrightarrow 0.
\end{equation}
As in the $N{=}1$ theory, this induces the long exact cohomology sequence
\begin{equation}
\longrightarrow H^{\pi^+,\pi^-}(\bar{F}) \stackrel{pr }{\longrightarrow}
H^{\pi^+ -1 ,\pi^-}(F_-) \stackrel{X_0^+}{\longrightarrow} H^{\pi^+,\pi^-}(F_-)
\stackrel{i}\longrightarrow H^{\pi^+, \pi^-}(\bar{F})  \longrightarrow.
\end{equation}
Using $H^{\pi^+,\pi^-}(\bar{F}) = 0$ we obtain the following

\begin{lemma}
The maps
\begin{equation}
X^+_0:\quad H^{\pi^+,\pi^-}(F_-)\ \longrightarrow\ H^{\pi^++1 ,\pi^-}(F_-)
\end{equation}
and 
\begin{equation}
X^-_0:\quad H^{\pi^+,\pi^-}(F_+)\ \longrightarrow\ H^{\pi^+ ,\pi^-+1}(F_+)
\end{equation}
are isomorphisms.
Thus $H^{\pi^+,\pi^-}(F_-)$ can depend only on $\pi^-$ 
and $H^{\pi^+,\pi^-}(F_+)$ only on $\pi^+$.
\end{lemma}
Note that this result holds for any value of the momentum. 

In the second step consider the projection
\begin{equation}
pr': F_- \to F,
\hspace{2cm}
pr (a + \xi_0^- b) = b. 
\end{equation}
Again it anticommutes with $Q$, and via the exact sequence
\begin{equation}
0 \longrightarrow F \stackrel{i}{\longrightarrow} F_-  
\stackrel{pr'}{\longrightarrow} F  \longrightarrow 0
\end{equation}
and the corresponding exact triangle one obtains for each pair $(\pi^+,\pi^-)$  
a long exact cohomology sequence with connecting homomorphism $X^-_0$:
\begin{eqnarray} \label{seq3}
\ldots\ \stackrel{X_0^-}{\longrightarrow}\ H^{g,\pi^+,\pi^-}(F)\
\stackrel{i}{\longrightarrow} & H^{g,\pi^+,\pi^-}(F_-) & \nonumber \\
& {\scriptstyle pr'} \downarrow & \\
& H^{g+1,\pi^+,\pi^- -1}(F) & \stackrel{X_0^-}{\longrightarrow}\ 
H^{g+1,\pi^+,\pi^-}(F)\ \stackrel{i}{\longrightarrow}\ \ldots. \nonumber
\end{eqnarray}

We first treat the exceptional case. In section 2.2 it has been shown that
\begin{equation}
H^{g,-1,0} (F|k{=}0)\ =\ H^{g,-1,-1} (F|k{=}0)\ =  0
\qquad {\rm for} \quad g \neq 1,2. 
\end{equation}
This can be inserted into (\ref{seq3}) at $\pi^+{=}-1$ and $\pi^-{=}0$ to yield
\begin{equation}
H^{g,-1,0}(F_-|k{=}0)\ = 0 \qquad {\rm for} \quad g \neq 0,1,2.
\end{equation}
Since spectral flow has ghost number zero, 
it follows from eqn.~(\ref{SFO1}) that
\begin{equation}
H^{g,\pi^+,\pi^-}(F_-|k{=}0)\ = 0 \qquad {\rm for} \quad
\pi^+{+}\pi^- = -1 \quad {\rm and} \quad g \neq 0,1,2.
\end{equation}
Lemma 1 guarantees that these cohomologies do not depend on $\pi^+$,
which forces them to vanish at {\it all\/} pictures. 
Once more using the sequence (\ref{seq3}) implies that $X_0^-$
is an isomorphism of the absolute exceptional cohomology for $g \neq 0,1,2,3$. 
An analogous argument proves that $X_0^+$ generates isomorphies likewise. Since
\begin{equation}
H^{g,-1,-1}(F|k{=}0)\ = 0 \qquad {\rm for} \quad g \neq 1,2
\end{equation}
we conclude that
\begin{equation}
H^{g,\pi^+,\pi^-}(F|k{=}0)\ = 0 \qquad {\rm for} \quad g \neq 0,1,2,3
\end{equation}
in an arbitrary picture. 
For $g\in\{0,1,2,3\}$ we have seen counterexamples in section~2.2.
Note that these results hold for the absolute cohomology only.

If $k{\cdot}k=0$ but $k \neq 0$ the situation is more difficult. We can, 
however, extract one more piece of information from the sequence (\ref{seq3}):

\begin{lemma}
If there exists a pair of  numbers $\hat{\pi}^+, \hat{\pi}^-$ such that
\begin{equation}
X^+_0:\quad H^{\hat{\pi}^+,\hat{\pi}^-}(F)\  
\longrightarrow\ H^{\hat{\pi}^+ +1,\hat{\pi}^-} (F) 
\end{equation}
is an isomorphism, then 
\begin{equation}
X^+_0:\quad H^{\hat{\pi}^+,\pi^-}(F)\  
\longrightarrow\ H^{\hat{\pi}^+ +1,\pi^-} (F) 
\end{equation}
is an isomorphism for arbitrary $\pi^-$. 
Spectral flow then establishes isomorphy of all $H^{\pi^+,\pi^-}(F)$.
An analogous result holds for $X^-_0$.
\end{lemma}
This statement is again valid for all momenta.

Lemma 2 is a direct application of the five-lemma 
from the theory of exact sequences (for example see ref.~\cite{S}).  
The  important observation is that the sequence (\ref{seq3}) depends on $\pi^+$.
To prove that $X^+_0$ is bijective we therefore write down the sequences for 
$\hat{\pi}^+$ and $\hat{\pi}^+ +1$ side by side and connect them by $X^+_0$
(due to lack of space the following diagram is rotated by $90^{\circ}$ 
from it usual form):
$$
\begin{CD}
H^{\hat{\pi}^+,\hat{\pi}^-+1}(F_-)   @>{X^+_0}>{\cong}>  
H^{\hat{\pi}^++1 ,\hat{\pi}^-+1}(F_-) \\
@V{pr'}VV               @V{pr'}VV    \\
H^{\hat{\pi}^+,\hat{\pi}^-}(F)  @>{X^+_0}>> 
H^{\hat{\pi}^+ +1,\hat{\pi}^- }(F) \\
@V{X_0^-}VV               @V{X_0^-}VV    \\
H^{\hat{\pi}^+,\hat{\pi}^- +1}(F)  @>{X^+_0}>> 
H^{\hat{\pi}^++1,\hat{\pi}^- +1}(F) \\
@V{i}VV               @V{i}VV    \\
H^{\hat{\pi}^+,\hat{\pi}^-+1}(F_-)  @>{X^+_0}>{\cong}> 
H^{\hat{\pi}^++1,\hat{\pi}^- +1}(F_-) \\
@V{pr'}VV               @V{pr'}VV    \\
H^{\hat{\pi}^+,\hat{\pi}^-}(F)  @>{X^+_0}>> 
H^{\hat{\pi}^++1,\hat{\pi}^- }(F) \\
\end{CD}
$$
Due to Lemma 1, the first and fourth horizontal maps  from the top 
are isomorphisms as indicated. Moreover, the columns are exact, and we have
\begin{equation}
[X^+_0 , pr'] = [X^+_0 , i] =0,
\hspace{1cm}
[X^+_0 , X^-_0] = X_0^+ \xi^-_0 Q - \xi^-_0 Q \xi^+_0 Q - Q[X^+_0,\xi^-_0].
\end{equation}
Since $[X^+_0,\xi^-_0]$ does not contain any $\xi^{\pm}_0$, 
this implies that all these maps commute on the cohomology spaces. 
Hence the diagram is commutative. 
If we now use the assumption of Lemma 2 that also the second and the bottom  
horizontal maps are isomorphisms, the five-lemma tells us that the horizontal 
map in the middle is an isomorphism, too.  
The lemma thus follows for all $\pi^- > \hat{\pi}^-$ by induction. 
The case $\pi^- < \hat{\pi}^-$ can be treated similarly. 
This concludes the proof.

Lemma 2 immediately applies to the massless case for non-zero momentum.
Indeed, the isomorphy in~(\ref{start}) now implies 
that $H^{\pi^+,\pi^-}(F|k{\neq}0)$ is picture independent.

We add that our proof extends to the massive case as well. 
To employ Lemma 2 however, one first needs to show that
$H^{-1,-1}(F)=H^{-1,0}(F)$, for example, at $k{\cdot}k>0$.
This gets involved due to the infinity of candidates states.

\section{Summary}
To be able to properly put into context the results of this paper 
we again describe the cohomology of the $N{=}1$ string. 
In $N{=}1$ string theory the absolute BRST cohomology is picture independent 
for any value of the momentum. This can be proven either by inverting 
the picture-raising operator in a momentum independent way~\cite{HMM} or 
by exploiting the fact that the absolute cohomology in the extended Fock space
is trivial~\cite{NQ}. At non-vanishing momentum this cohomology contains two 
copies of the space of states obtained e.g. by light-cone quantisation. 
At zero momentum even more states appear. To obtain a one-to-one relation 
between BRST and light-cone quantisation, it is necessary in addition to impose
on physical states the condition $b_0 |phys \rangle = 0$ which leads to the 
more relevant relative cohomology. 
Unfortunately, the analysis of refs.~\cite{HMM, NQ} does not apply to the 
relative cohomology. However, it has been shown recently~\cite{BZ} 
that the picture-raising operator can be inverted on states of 
non-vanishing momentum by an operator that commutes with $b_0$, thereby proving
picture independence of the relative cohomology at non-vanishing momentum. 
At zero momentum the above argument does not work and, 
besides the fact that the zero-momentum cohomology is generally larger 
than the zero-momentum limit of the non-zero-momentum cohomology, there 
also is a picture dependence in the relative (but not in the absolute) case 
as has been demonstrated explicitly~\cite{BZ}.

In $N{=}2$ string theory, the relative BRST cohomology 
in the $({-}1,{-}1)$ picture can be computed rigorously 
along the lines of refs.~\cite{FGZ, LZ}, as we desribed in section 2.3. 
There exists only a single massless physical state at ghost number one. 
However, the issue of picture independence of the BRST cohomology 
has long been unclear since the picture-raising operators cannot be inverted 
in a momentum independent way. Hence, an argument analogous to that 
of ref.~\cite{HMM} for the $N{=}1$ string does not exist in this theory. 
Nevertheless, we proved the picture independence of the relative 
and the absolute cohomology at non-vanishing momentum,
providing two different methods for the absolute massless case.

In section 3 we applied the ideas of ref.~\cite{BZ} to the $N{=}2$ string 
thus showing the picture independence at non-vanishing momentum 
in a rather straightforward way. 
In section 4 we gave an alternative treatment based on the fact that, as in the
$N{=}1$ theory, the absolute cohomology in the extended Fock space is trivial. 
Combined with the spectral flow automorphism of the $N{=}2$ superconformal 
algebra and explicit computations in simple pictures we again proved inductively
that picture raising is an isomorphism of the absolute massless cohomology at 
non-zero momentum, without refering to any kind of picture-lowering. However, 
this argument is restricted to the absolute cohomology and does not apply 
to the relative case. Higher mass levels can principally be treated in the
same way. 

The virtue of the rather complicated analysis of section 4 is that, 
in contrast to the argument of section 3, it also allows one to constrain the 
exceptional cohomology at zero momentum where the concept of picture-lowering 
breaks down completely. By explicit computation in this case, we found a picture
dependence of both the relative and the absolute exceptional cohomology. 
In the $({-}1,{-}1)$ picture we showed that the zero-momentum cohomology 
is simply the zero-momentum limit of the cohomology at non-vanishing momentum. 
In the $({-}1,0)$ picture, however, 
the relative cohomology is two-dimensional at any positive ghost number, 
whereas the absolute cohomology is two-dimensional for ghost numbers 
one and two only. In higher pictures physical states with ghost numbers 
zero and three also appear.
About the zero-momentum case we could only prove that
its absolute cohomology vanishes at any picture for ghost numbers 
$g \neq 0,1,2$ or $3$. The picture dependence does not disappear
when taking into account the center-of-mass coordinate of the string. 
The computed dimensions of the zero-momentum cohomologies are summarised
as follows, with $\pi:=\pi^++\pi^-$.

\vspace{0.5cm}
\begin{tabular}{|r|cccccc|cccccc|}
\hline
$g{\scriptstyle\searrow}$ & & \multicolumn{4}{c}{dim $H(F)$} & & & 
\multicolumn{4}{c}{dim $H(F_{rel})$} & \\
$\pi\downarrow$ &
$<0$ & 0 & 1 & 2 & 3 & $>3$ & $<0$ & 0 & 1 & 2 & 3 & $>3$ \\
\hline
-4 & 0 &   &   & 3 & 1 & 0 &   &   & 2 & 1 & 0 & 0 \\ 
-3 & 0 & 0 & 2 & 2 & 0 & 0 & 2 & 2 & 2 & 0 & 0 & 0 \\
-2 & 0 & 0 & 1 & 1 & 0 & 0 & 0 & 0 & 1 & 0 & 0 & 0 \\
-1 & 0 & 0 & 2 & 2 & 0 & 0 & 0 & 0 & 2 & 2 & 2 & 2 \\
 0 & 0 & 1 & 3 &   &   & 0 & 0 & 1 & 2 &   &   &   \\
\hline
\end{tabular}
\vspace{0.5cm}

\noindent
The existence of a non-degenerate scalar product on the full cohomology
implies a pairing
\begin{eqnarray}
(g\ ,\ \pi^+\ ,\ \pi^-)\ &\longleftrightarrow &\ (3-g\ ,\ -2-\pi^+\ ,\ -2-\pi^-)
\qquad {\rm for} \quad F \nonumber \\
(g\ ,\ \pi^+\ ,\ \pi^-)\ &\longleftrightarrow &\ (2-g\ ,\ -2-\pi^+\ ,\ -2-\pi^-)
\qquad {\rm for} \quad F_{rel}
\end{eqnarray}
so that the dimensions of the corresponding cohomologies coincide.

We know from the BRST quantisation of gauge theories that extra physical states
at zero momentum are remnants of gauge and ghost degrees of freedom off the
mass shell. These are necessary for a covariant formulation but disappear
when fixing a gauge and going on-shell. In string theory, such states should
signal gauge symmetries present in a covariant {\it string field\/} formulation.
The observed picture dependence then suggests a proliferation of field degrees 
of freedom in higher pictures, in tune with the results of~\cite{DL}.
Work in this direction is in progress.

\vspace{1cm}

\noindent
{\bf Acknowledgements}

\noindent
K.J. would like to thank P. Adamietz and J. Fuchs for useful discussions.

\vfill\eject

\end{document}